\documentclass[10pt, sigconf,letterpaper]{acmart}
\usepackage{xspace}
\usepackage{subcaption}
\usepackage{tabularx}
\usepackage{makecell}

\AtBeginDocument{%
  \providecommand\BibTeX{{%
    \normalfont B\kern-0.5em{\scshape i\kern-0.25em b}\kern-0.8em\TeX}}}

\begin{document}

\title{Loss-tolerant neural video codec aware congestion control for real time video communication}

\author{Zhengxu Xia, Hanchen Li, Junchen Jiang}
\affiliation{%
  \institution{University of Chicago}
}




\newcommand{\name}{\textsc{NVC-CC}\xspace}

\newcommand{\zx}[1]{{\color{violet}{\footnotesize [ZX: #1]\xspace}}}

\newcounter{packednmbr}
\newenvironment{packedenumerate}{\begin{list}{\thepackednmbr.}{\usecounter{packednmbr}\setlength{\itemsep}{0.5pt}\addtolength{\labelwidth}{-4pt}\setlength{\leftmargin}{2ex}\setlength{\listparindent}{\parindent}\setlength{\parsep}{1pt}\setlength{\topsep}{0pt}}}{\end{list}}
\newenvironment{packeditemize}{\begin{list}{$\bullet$}{\setlength{\itemsep}{0.5pt}\addtolength{\labelwidth}{-4pt}\setlength{\leftmargin}{2ex}\setlength{\listparindent}{\parindent}\setlength{\parsep}{1pt}\setlength{\topsep}{2pt}}}{\end{list}}
\newenvironment{packedpackeditemize}{\begin{list}{$\bullet$}{\setlength{\itemsep}{0.5pt}\addtolength{\labelwidth}{-4pt}\setlength{\leftmargin}{\labelwidth}\setlength{\listparindent}{\parindent}\setlength{\parsep}{1pt}\setlength{\topsep}{0pt}}}{\end{list}}
\newenvironment{packedtrivlist}{\begin{list}{\setlength{\itemsep}{0.2pt}\addtolength{\labelwidth}{-4pt}\setlength{\leftmargin}{\labelwidth}\setlength{\listparindent}{\parindent}\setlength{\parsep}{1pt}\setlength{\topsep}{0pt}}}{\end{list}}
\let\enumerate\packedenumerate
\let\endenumerate\endpackedenumerate
\let\itemize\packeditemize
\let\enditemize\endpackeditemize

\newcommand{\tightcaption}[1]{
\vspace{-0.2cm}
\caption{{\normalfont{\textit{{#1}}}}}
\vspace{-0.6cm}
}
\newcommand{\tightsection}[1]{\vspace{-0.2cm}\section{#1}\vspace{-0.1cm}}
\newcommand{\tightsectionstar}[1]{\vspace{-0.17cm}\section*{#1}\vspace{-0.08cm}}
\newcommand{\tightsubsection}[1]{\vspace{-0.25cm}\subsection{#1}\vspace{-0.1cm}}
\newcommand{\tightsubsubsection}[1]{\vspace{-0.01in}\subsubsection{#1}\vspace{-0.01cm}}

\newcommand{\eg}{{\it e.g.,}\xspace}
\newcommand{\ie}{{\it i.e.,}\xspace}
\newcommand{\etal}{{\it et.~al}\xspace}
\newcommand{\bigO}{\mathrm{O}}
\newcommand{\twlog}{w.l.o.g.\xspace}

\newcommand{\myparashort}[1]{\vspace{0.05cm}\noindent{\bf {#1}}~}
\newcommand{\mypara}[1]{\vspace{0.05cm}\noindent{\bf {#1}:}~}
\newcommand{\myparatight}[1]{\vspace{0.02cm}\noindent{\bf {#1}:}~}
\newcommand{\myparaq}[1]{\smallskip\noindent{\bf {#1}?}~}
\newcommand{\myparaittight}[1]{\smallskip\noindent{\emph {#1}:}~}
\newcommand{\question}[1]{\smallskip\noindent{\emph{Q:~#1}}\smallskip}
\newcommand{\myparaqtight}[1]{\smallskip\noindent{\bf {#1}}~}

\newcommand{\fillme}{{\bf XXX}\xspace}


\begin{abstract}

Because of reinforcement learning's (RL) ability to automatically create more adaptive controlling logics beyond the hand-crafted heuristics, numerous effort has been made to apply RL to congestion control (CC) design for real time video communication (RTC) applications and has successfully shown promising benefits over the rule-based RTC CCs. Online reinforcement learning is often adopted to train the RL models so the models can directly adapt to real network environments. However, its trail-and-error manner can also cause catastrophic degradation of the quality of experience (QoE) of RTC application at run time. Thus, safeguard strategies such as falling back to hand-crafted heuristics can be used to run along with RL models to guarantee the actions explored in the training sensible, despite that these safeguard strategies interrupt the learning process and make it more challenging to discover optimal RL policies.

The recent emergence of loss-tolerant neural video codecs (NVC) naturally provides a layer of protection for the online learning of RL-based congestion control because of its resilience to packet losses, but such packet loss resilience have not been fully exploited in prior works yet. In this paper, we present a reinforcement learning (RL) based congestion control which can be aware of and takes advantage of packet loss tolerance characteristic of NVCs via reward in online RL learning. Through extensive evaluation on various videos and network traces in a simulated environment, we demonstrate that our NVC-aware CC running with the loss-tolerant NVC reduces the training time by 41\% compared to other prior RL-based CCs. It also boosts the mean video quality by 0.3 to 1.6dB\%, lower the tail frame delay by 3 to 200ms, and reduces the video stalls by 20\% to 77\% in comparison with other baseline RTC CCs.
\end{abstract}





\setcopyright{none}
\renewcommand\footnotetextcopyrightpermission[1]{}
\thispagestyle{plain}
\pagestyle{plain}

\maketitle


\section{Introduction}
\label{sec:intro}

Real-time video communication including video conferencing~\cite{macmillan2021measuring}, live video/VR broadcasting
~\cite{WebrtcVRStreaming, WebRTCARVR, hopkins2017live}, IoT applications~\cite{WebRTCIoT, MobidevWebrtcIoT}, and cloud gaming~\cite{WebrtcStadia, WebrtcCloudGaming} has been a key component of our daily lives~\cite{blum2021webrtc} and carries a dominant amount of traffic in today's internet~\cite{CiscoAnnualInternetReport}.

These RTC applications require high network bandwidth and low network latency to deliver seamless and high quality experience to users, pushing the telecommunication infrastructure upgrade to meet the demands and forcing the congestion control algorithms to promptly adapt to the constantly changing network conditions.
Unlike traditional congestion controls~\cite{tcpcubic, tcpvegas, tcpcc} which are designed for reliability and in-order delivery through retransmissions instead of realtimeliness, plenty of hand-crafted congestion controls for real-time video communication ~\cite{GCC, Salsify, NADA, SCReAM, FBRA, sqp} have been proposed to boost bandwidth utilization, suppress packet delays, and avoid packet losses. However, these pre-programmed rule-based congestion control algorithms are not panacea in all network settings as they fall short of adapting to the highly heterogeneous network conditions.

To save the human effort optimizing a rule-based congestion control algorithm for numerous network conditions, researchers have made huge effort to explore the data-driven approaches to design congestion control and rate adaptation~\cite{Pensieve, Aurora, OnRL, Loki, Concerto, Genet, robustifying} and have shown great potential over the handcrafted heuristics.
The other approach designed to bridge the gap between training network environments and the real networks is using
The ``learning online, running online" strategy is often adopted to train a RL-based solution in order to bridge the gap between training network environments and the real network environments at the deployment stage. RL models directly interact with the real network environments to collect experience and then update themselves during runtime. However, the trial-and-error behavior of online RL training will unavoidably take risky actions which might disturb the system performance. A safeguard policy, typically a handcrafted heuristics, is often used to substitute the RL model once an erroneous action is detected or the system is in a risky state~\cite{OnRL, mao2019towards}. After the safeguard policy recovers the system to a safe state, the RL-based model takes the control back.

\begin{figure}[t]
    \centering
    \includegraphics[width=0.73\columnwidth]{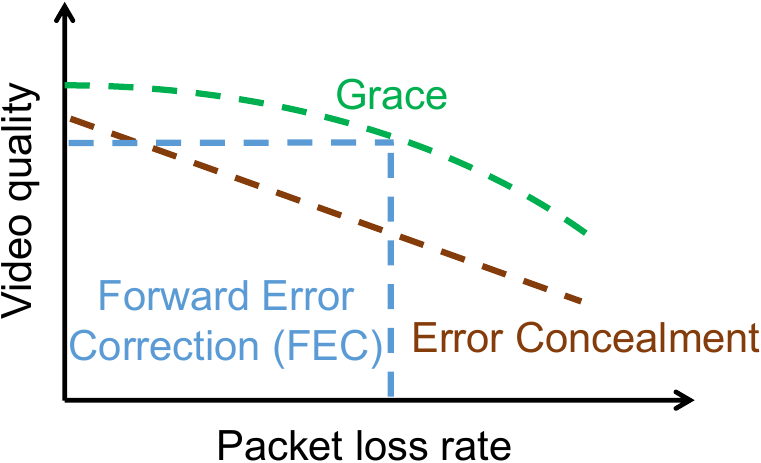}
    \tightcaption{\bf Loss-tolerant neural video codec has slow and smooth video quality drop with increasing packet loss rate.}
    \label{fig:video-quality-of-loss-resilience-schemes}
\end{figure}

The main design philosophy behind safeguarding a RL-based CC as well as behind traditional CCs in RTC applications is based on an implicit assumption on video codec that delayed or lost packets can lead to incomplete frames received which then block video decoding at the receiver side and hurt users' QoE.
The recent loss-tolerant neural video codecs~\cite{Swift, FVC, DVC, Grace, Lifter, Gemino} breaks the implicit assumption on video codecs as these NVCs can decode incomplete frames and still deliver decent frame quality. They have shown strong loss tolerance ability across a wide range of packet loss rates on top of its high compression efficiency and good generalization over various video content. Figure~\ref{fig:video-quality-of-loss-resilience-schemes} shows a state-of-art neural video codec, GRACE, has a smoother and slower video quality drop with increasing packet loss rate than commonly used encoder-side forward error correction (FEC) and decoder-side error concealment (EC).

In this paper, we present \name, a RL-based RTC congestion control algorithm which can be trained online without the help of safeguard policies by taking advantage of the NVCs' packet loss tolerance properties. Our key insight is that \textit{the safeguard policies run along the RL model hinders the RL online learning efficiency}.

Comprehensive experiments (\S\ref{sec:eval}) on a diverse set of videos and network traces show that our NVC-aware CC running with the loss-tolerant NVC reduces the training time by 41\% compared to other prior RL-based CCs. It also boosts the mean video quality by 0.3 to 1.6dB\%, lower the tail frame delay by 3 to 200ms, and reduces the video stalls by 20\% to 77\% in comparison with other baseline RTC CCs.

\noindent{\bf Contributions:} Our work makes the following contributions.

\noindent{1)} We reveal the inefficiency of training in RL-based RTC congestion control solutions trained by online learning with safeguard policies and introduce the \textit{trade-off between learning efficiency and QoE in RL training} (\S\ref{subsec:mot-inefficient-training}).

\noindent{2)} We analyze how the loss-tolerant NVCs can help improve training efficiency by allowing RL-based CCs to learn without the restriction of safeguard policies and not not hurting QoE (\S\ref{subsec:nvc-help}).

\noindent{3)} We propose \name, which, to the best of our knowledge, the first RL-based congestion control aware of and taking advantage of the loss-resilient properties of neural video codecs (\S\ref{sec:design}) and validate its remarkable performance gain over the state-of-the-art solutions (\S\ref{sec:eval}).




\section{Background}
\label{sec:background}

To help explain \name 's design, we first introduce some important concepts in real-time video communication.

\subsection {Real-time video communication}
\begin{figure}[t]
    \centering
    \includegraphics[width=0.82\columnwidth]{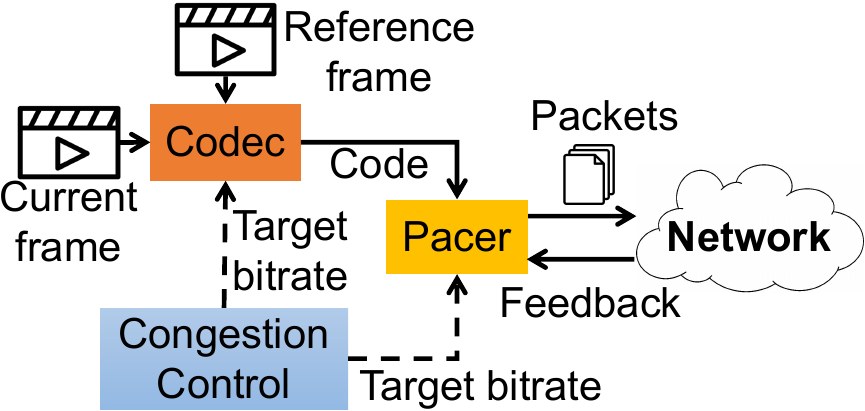}
    \tightcaption{\bf RTC workflow.}
    \label{fig:real-time-video-communication-workflow}
\end{figure}

Figure \ref{fig:real-time-video-communication-workflow} shows a typical RTC workflow. Given a new frame and a reference frame,
1) the video codec at the sender encodes the new frame based on a target encoding bitrate estimated by the congestion control module;
2) the sender's pacer packetizes the code and paces the packets into the network according to a target sending bitrate estimated by the congestion control module;
3) the congestion control module simultaneously collects the feedback from the network and adapts the target bitrate continuously; and
4) finally the video codec at the receiver decodes the received data and reconstruct each frame.


Because lost or heavily delayed packets can affect frame decoding and introduce undesired frame quality drop, frame delay and video stalls to end users, RTC CCs need to deal with available bandwidth fluctuations with the best effort. However, it is impossible for the CC module to perfectly predict the future bandwidth changes, making heavily delayed packets and packet losses unavoidable. Thus, loss-resilient techniques and video codecs play a vital role in frame decoding especially under packet losses.

To differentiate the packet loss in network level, we define \textit{packet loss per frame} as any packets not received before the receiver is expected to decode the frame~\cite{Grace}.

\subsection{Neural video codec} 
\begin{figure}[t]
    \centering
    \includegraphics[width=0.96\columnwidth]{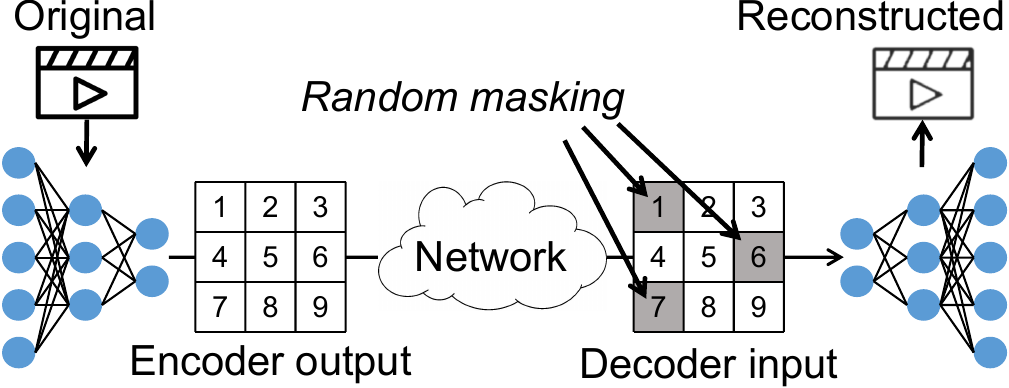}
    \tightcaption{\bf GRACE's training procedure.}
    \label{fig:grace-training}
\end{figure}

Neural video codec (NVC) consists of trained neural networks instead of handcrafted logic as video encoder and decoder. Prior works in NVCs~\cite{Swift, FVC, DVC} have demonstrated comparable or even better compression efficiency comparing to traditional video codecs like H.265~\cite{HEVC}, and VP9~\cite{VP9} because they replace handcrafted heuristics in the common logical components of traditional video codes, such as motion estimation, warping, and transformative compression, with neural networks, which can learn more complex algorithms from data. Prior NVCs have also shown great generalization across a variety of video content due to their ability to be trained on a huge amount of videos. Additionally, recent studies~\cite{Grace, Lifter, Gemino} have discovered that NVCs have stronger loss resilience over a wider range of packet loss rates comparing to traditional loss resilience schemes (shown in Figure ~\ref{fig:video-quality-of-loss-resilience-schemes}). To obtain stronger loss tolerance ability, the neural encoder and decoder are jointly optimized via training directly across packet loss rates. Figure \ref{fig:grace-training} illustrates the training procedure of the loss-tolerant NVC, GRACE. Unlike traditional NVC training that assumes no data loss between the encoder and decoder, GRACE applies ``random masking''--setting a fraction of randomly selected elements to zeros--to the encoder’s output to simulate network packet losses. However, an important missing piece missing from prior works is that how should congestion control leverage these loss tolerant NVCs to yield better QoE.

%
%

\section{Motivation}
\label{sec:motivation}
In this section, we will first verify the effectiveness of RL-based solutions in the field of RTC communication. Then, we are going to take OnRL~\cite{OnRL} as an example to show that the existence of safeguard can cause learning efficiency problem and analyze the tradeoff between learning efficiency and the QoE during RL training. Lastly, we motivate how a loss-tolerant NVC can mitigate this problem.

\subsection{RL-based CCs are promising}
\label{subsec:mot-rl-promising}
Recently proposed RL-based CCs have impressive improvement over traditional rule-based CCs. The RL-based CC, usually implemented by a neural network (NN), works with the transport layer and video codecs to collect states from the network environment. It then makes a decision on sending rate and broadcast the latest sending rate to the transport layer and video codes. During the training stage, its goal is to maximize a reward (typically a combination of throughput, latency, and packet loss rate).

The reasons of adopting RL in RTC CC are two-fold. Firstly, learning-based congestion controls have significant potential to self-adapt to different network conditions, eliminating the necessity for manual tuning or engineering for each specific network scenario. Contrast to traditional rule-based CCs like GCC~\cite{GCC} which have taken engineers years of effort to optimize to various network environments, the learning-based approach allows the adaptation to new network environments to be automatically done by machines in hours or days.

Secondly, CC problem is a sequential decision-making process, which fits well in the scope of reinforcement learning. Additionally, the neural network in deep reinforcement learning (DRL) exposes the potential and ability to learn from raw input data without the need of tedious data preprocessing and handcrafted feature engineering as long as the training network environment distribution align with the testing distribution.

\begin{figure}[t]
    \centering
    \includegraphics[width=0.7\columnwidth]{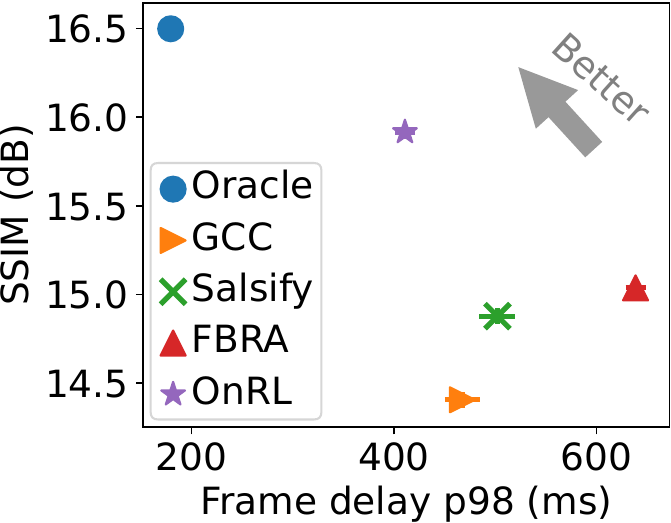}
    \tightcaption{\bf RL-based CC has shown prospects over traditional rule-based CCs.}
    \label{fig:rl-promising}
    \vspace{8pt}
\end{figure}

We have conducted simulation experiments to compare a RL-based solution, OnRL, to several well-known rule-based CCs and the results do verify prior works' findings that RL-based solutions can outperform traditionally rule-based approaches. We train and compare OnRL against rule-based CCs with a traditional video codec (e.g. H.264) on 2850 video sessions (50 diverse synthetic network traces randomly drawn from the network environment distribution in Table~\ref{tab:trace-parameters} and 57 videos from Table~\ref{tab:video-datasets}). Figure~\ref{fig:rl-promising} plots the average frame quality and average tail frame delay tradeoff of different CCs and it does show that OnRL is better than the rule-based approaches over the network environments in the same distribution that it trains on.


\subsection{Inefficient online RL training with safeguards}
\label{subsec:mot-inefficient-training}
RL-based solutions are not perfect. To obtain such a well-trained OnRL in Figure~\ref{fig:rl-promising} requires itself to be trained across 1200 networking environments drawn in the same distribution with a total duration of 10 hours. Given that RL models learn by exploring the action space and collecting the environment feedback as mentioned in \S\ref{sec:intro}, they will inevitably conduct risky actions which  may overwhelm the network capacity to cause packet delay increase or packet losses or underutilize the network available bandwidth. These risky actions eventually incur huge frame delay, poor video quality, and video stalls hurt QoE because traditional video codecs cannot decode incomplete frames and further packet retransmission might be needed.

To prevent such risky actions in RL training (especially online RL training), prior works~\cite{OnRL, mao2019towards} detect if risky actions are made and immediately fall back to a rule-based policy to recover the system to a safe state. Only after the system is not in a risky state anymore, the RL policy mode is switched back. For example, OnRL~\cite{OnRL} falls back to its safeguard policy, GCC, when its RL model action leads to an obvious packet jitter greater than the dynamically adjusted threshold. On the action which triggers a policy switch, it is heavily penalized so that the model is expected to learn to avoid the policy switch in the future. However, the existence of safeguard policies prevent the RL model from learning efficiently~\cite{gu2024enhancing}.

\begin{figure}[t]
    \centering
    \includegraphics[width=0.77\columnwidth]{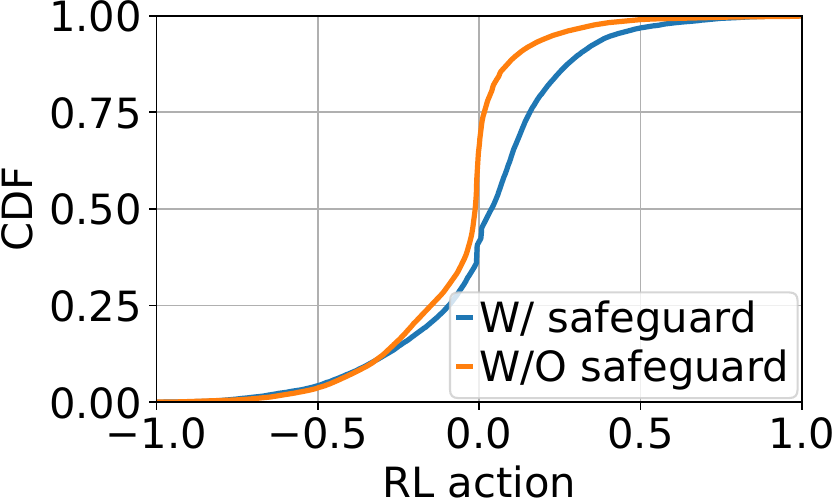}
    \tightcaption{\bf Comparison between CDFs of sampled actions from RLs trained with and without the safeguard policy.}
    \label{fig:action-space-distribution-comparison}
\end{figure}



The reasons are threefold. Firstly, the existence of safeguard policies can prevent action space from being explored thoroughly and diversely during training. We verify that safeguard policies limit RL action space exploration by comparing of the distributions of sampled actions between training the RL models with and without the safeguard policy respectively on an identical network environment. Figure~\ref{fig:action-space-distribution-comparison} shows that OnRL model with the safeguard policy spends more than 70\% of time exploring the region of action space which increases the sending rates while the training without the safeguard policy balance its sampling in both action space regions that increase sending rate and regions that decrease sending rate. The diversity of actions selected by RL model with safeguard is thus much narrower than that without safeguard and RL model may potentially miss better action sequences, which in turn slow down the learning efficiency.

Secondly, the existence of safeguard policies limit RL observation space exploration. An intuitive example is to verify that safeguard policies limit RL observation space exploration is that an OnRL model trained in a network environment with a deep queue network is brought to a network environment with a very short queue. Packets are dropped without apparent latency increase so no safeguard will be triggered. Because RL model never sees network observations with packet losses in previous training, it may conduct risky actions and then cause the QoE degradation until it learns from the network observations with packet loss.

\begin{figure}[t]
    \centering
    \includegraphics[width=0.88\columnwidth]{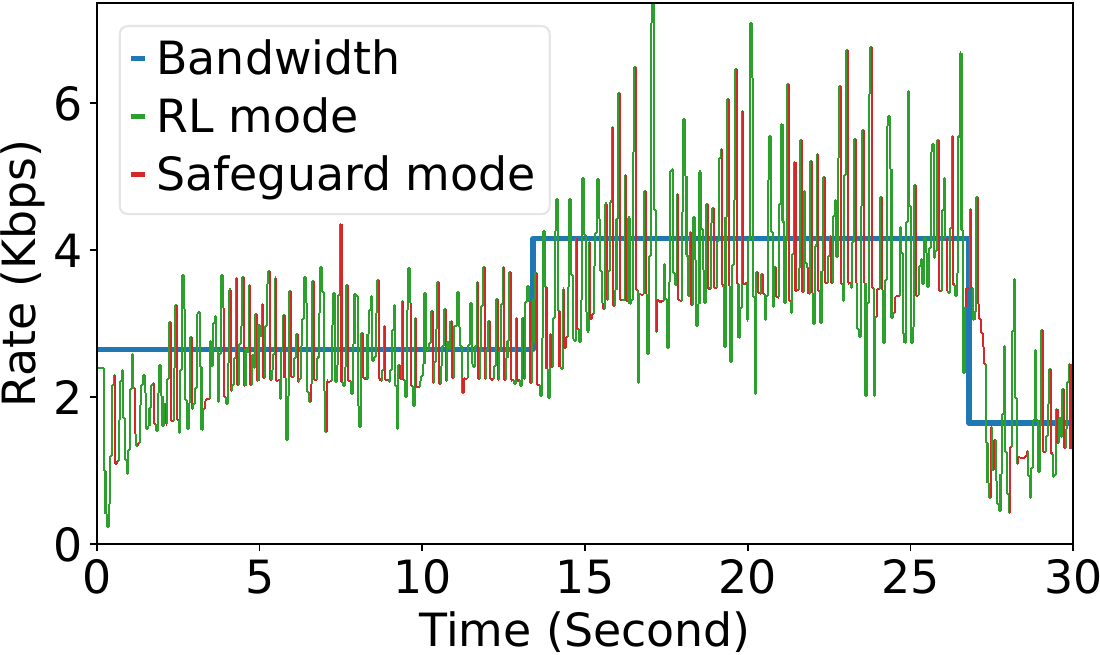}
    \tightcaption{\bf OnRL triggers GCC-based safeguard very frequently causing low RL training sample efficiency. The network has one-way link propagation delay of 25ms, a FIFO queue with capacity of 30KB and no random loss.}
    \label{fig:onrl-mode-switch}
\end{figure}

Thirdly, the sampling efficiency when training RL model with safeguard policies is low. The trajectory (the sequence of tuples formed by actions, rewards, and network observations) collected by the mixture of RL model and the safeguard policy cannot be directly passed to the NN training optimizer because it violates the assumption that on-policy RL learning algorithms like PPO~\cite{ppo} and Markov Decision Process modeling of RL require that every action should be from the RL model. We also empirically show that safeguard in OnRL is triggered too frequently causing low sampling efficiency during the training process.
For instance, if OnRL is trained on a single network environment with respect to a 5min video and the snapshot of the first 30 seconds, Figure~\ref{fig:onrl-mode-switch} shows that OnRL switches into GCC for 160 times and stays in GCC for 8 seconds which are wasted in OnRL training within 30 seconds.

\begin{figure}[t]
    \centering
    \includegraphics[width=0.80\columnwidth]{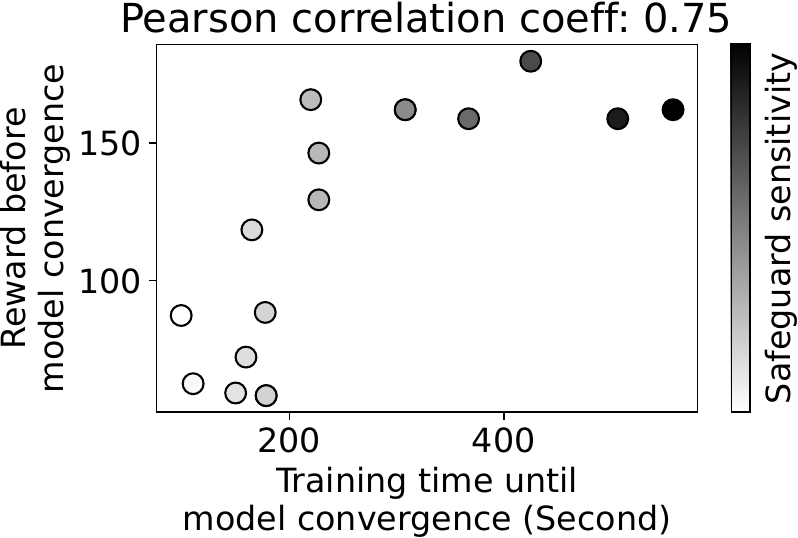}
    \tightcaption{\bf The tradeoff between QoE (training reward) and the time RL training convergence needs. The safeguard is triggered more frequently as the color (safeguard sensitivity) darkens.}
    \label{fig:quality-degradation-learning-efficiency-tradeoff}
\end{figure}

Therefore, there naturally exists a tradeoff between the learning efficiency of an RL model to convergence and the reward or QoE in the RL training stage. We define the learning efficiency as the number of seconds used to train an RL model to converge on a network environment. The model convergence is defined such that the RL model's testing reward on a training network environment does not vary by more than 10\% if the training on this network environment continues. Following the methodology in OnRL~\cite{OnRL}, we use RL reward in Equation~\ref{eq:aurora-reward} to represent video QoE. We verify the tradeoff by training RL with and without its safeguard on a single environment and plot the training reward before model convergence and the training time used until model convergence in Figure~\ref{fig:quality-degradation-learning-efficiency-tradeoff}. We sweep a wide range of safeguard sensitivity to demonstrate different degree of safeguard involvement in RL-based CC.

\subsection{How can a loss-tolerant NVC help RL-based CC training?}
\label{subsec:nvc-help}
Before talking about how a loss-tolerant NVC can help RL-based CC training, we first discuss its differences comparing to a traditional video codec. The differences lie in the following two aspects:

\mypara{Difference 1} NVC can translate packet delays and packet losses into video frame quality drop much smoother than a traditional video codec because the NVC can decode an incomplete frame which is not decodable for traditional video codecs as shown in Figure~\ref{fig:video-quality-of-loss-resilience-schemes}. A takeaway is that a risky action from RL-based CC is not risky any more if the video codec is changed from traditional ones to NVC and RL-based CC can explore more actions than before.

\begin{figure}[t]
    \centering
    \includegraphics[width=0.90\columnwidth]{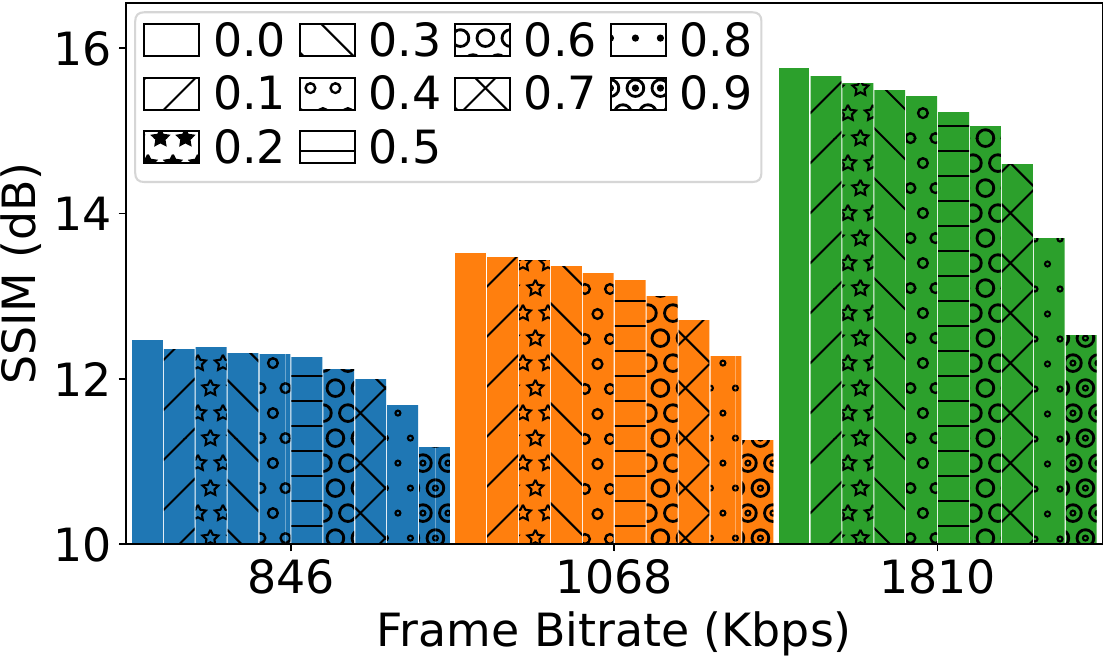}
    \tightcaption{\bf GRACE frame quality under three encoding bitrates and various frame-level packet loss rates. Bars with the same color but different hatches represent the quality of a frame encoded at the same bitrate and received with different frame loss rates.}
    \label{fig:grace-frame-quality-bar-plot}
    \vspace{8pt}
\end{figure}

\mypara{Difference 2} A high-bitrate video frame with minor packet loss rates (\ie less than 10\%) encoded by NVC can have better quality than a low-bitrate video frame with zero packet loss does. Figure~\ref{fig:grace-frame-quality-bar-plot} plots the qualities of a video frame encoded by GRACE under different bitrates and frame-level packet loss rates given that the reference frame is received completely. The frame quality encoded at 1810Kbps with 10\% frame data lost is approximately 2dB better that encoded at 1068Kbps with no frame data lost. However, because quality degradation caused by incomplete reference frames will propagate along frames. \cite{Grace} suggests that a NVC state synchronization is required every 10 frames when contiguous frame losses happen. But in between two NVC state synchronization, a takeaway is that sending frames at high-bitrate with minor packet loss at frame level may yield better frame quality.

\begin{figure}[t]
    \centering

    \begin{subfigure}[t]{0.49\linewidth}
        \centering
        \includegraphics[width=\linewidth]{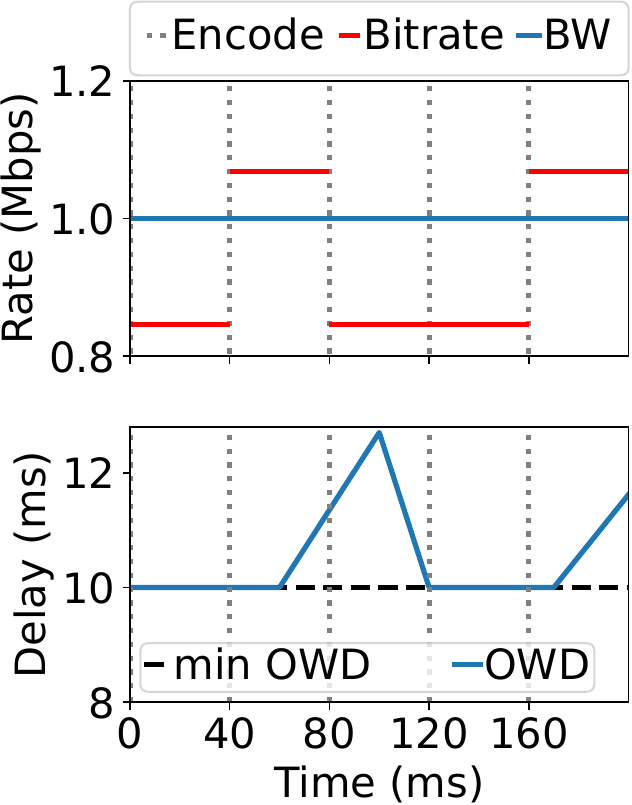}
        \caption{With safeguard}
        \label{fig:action-compare-with-safeguard}
    \end{subfigure}
    ~~~
    \begin{subfigure}[t]{0.49\linewidth}
        \centering
        \includegraphics[width=\linewidth]{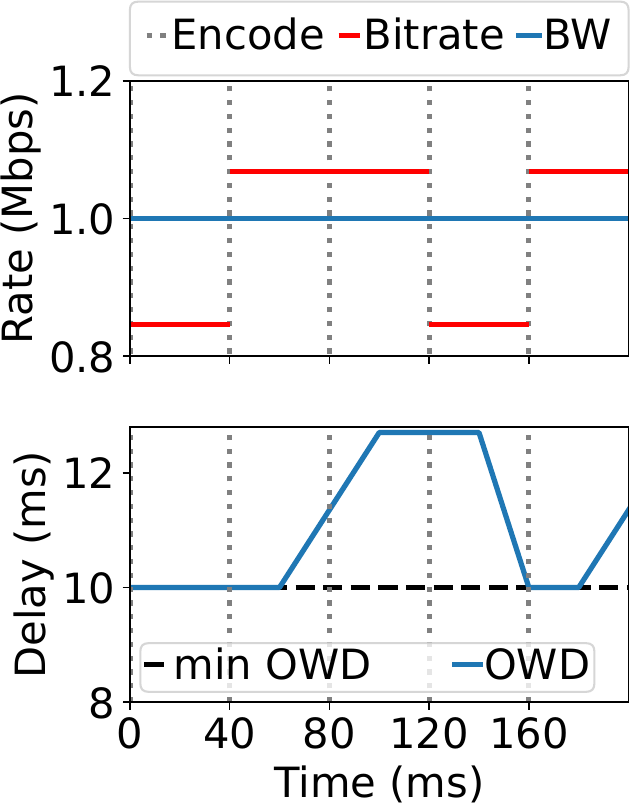}
        \caption{Without safeguard}
        \label{fig:action-compare-without-safeguard}
    \end{subfigure}
    \tightcaption{\bf Action sequence comparison between RLs training with or without a safeguard policy. The network environment has constant bandwidth of 1.0Mbps, 10ms minimum one-way delay (OWD), and a very shallow queue of 1.5KB.}
    \label{fig:action-compare}
\end{figure}

Figure~\ref{fig:action-compare} shows an example that the safeguard policy prevents the RL model from finding a better sequence of actions when the RL-based CC is trained with the loss-tolerant NVC, GRACE. In this figure, CCs is assumed to make a bitrate decision on every frame encoding event (every 40ms), and the frame data are smoothly paced out into the network before the next frame comes. As illustrated in Figure~\ref{fig:action-compare-with-safeguard}, every time the RL chooses a bitrate overshooting the bandwidth and causes the delay inflation, the safeguard immediately takes over the control and reduces the bitrate. A possible sequence of actions that can be explored by RL without the safeguard whereas RL with safeguard will never explore due to the existence of the safeguard is in Figure~\ref{fig:action-compare-without-safeguard}. By enduring the tail delay 12.7ms and less than 50\% frame loss rate, the frame sent at 80ms is approximately 1dB in Figure~\ref{fig:action-compare-without-safeguard} higher than that in Figure~\ref{fig:action-compare-without-safeguard}. The average frame quality over 5 frames is about 0.2dB better.

\section{Design}
\label{sec:design}
So far we have shown that the RL-based RTC CCs look promising over traditional rule-based RTC CCs and how a loss-tolerant NVC can help the RL training if RL-based CC can be aware of it. We thus propose a new RL-based RTC CC design which leverages the loss tolerance characteristic of recent loss-tolerant NVCs via changing the reward function design.



\mypara{State, action, and NN architecture}
\name's observed states are designed based on those of Aurora~\cite{Aurora} and OnRL~\cite{OnRL}. A observed state is collected every time 50ms window, where the window length is equal to the RTCP feedback interval. It consists of the following: (i) latency gradient, the derivative of packet latency with respect to time; (ii) latency ratio, the ratio of the current time window's mean packet latency to the minimum observed mean packet latency of any time windows in the connection's history; (iii) sending ratio, the ratio of packets sent to packets acknowledged by the receiver. To resolve the oscillation problem of sending rates of RL decision encountered in prior works, the action space is not a discrete space like OnRL's but a continuous space $[-1, 1]$. \name takes an input feature vector constructed by the states of 10 time windows and makes an action $a_t$ to adjust the sending rate $x_t$ for the next time window $t$ according to Equation~\ref{eq:rl-action}. $\gamma$ is the throughput measured within the past 200ms. The NN architecture of \name is a small fully connected neural network with two hidden layers composed of 32 and 16 neurons respectively and tanh nonlinearity.
\begin{equation}
    x_t =
        \begin{cases}
            x_{t-1} * a_t & a_t > 0\\
            \gamma / a_t & a_t < 0
        \end{cases}
    \label{eq:rl-action}
\end{equation}
\mypara{Reward function}
Existing RL-based CCs~\cite{Aurora, Loki, OnRL, Genet} typically compute the linear combination of throughput, packet delays, packet losses as the reward instead of directly using the feedback of video codecs because they assume a traditional video codec runs on top and cannot reflect the state of the network smoothly. For example, Aurora~\cite{Aurora} observes network states in a time window to optimize for a reward as shown in Equation~\ref{eq:aurora-reward}. Throughput (Tput) in kbps, latency (Lat) in second, and loss rate (Loss) are measured in a time window and $a=120$, $b=-1000$, $c=-2000$.
\begin{equation}
    R=\sum_i \frac{1}{n}(a\cdot \text{Tput}_i + b\cdot \text{Lat}_i + c\cdot \text{Loss}_i)
    \label{eq:aurora-reward}
\end{equation}
Instead of indirectly optimizing QoE of real-time communication session via a reward based on network-level performance, we propose to optimize a reward directly on frame quality and packet delays, enabled by the loss-tolerant video codecs. One may argue that QoE of traditional video codecs can also be used in RL-based CC training. However, because traditional video codecs cannot always decode a frame and reflect the frame quality and delay to the sender within every decision step especially when there are packet losses or heavy packet queuing. The trial and error manner in RL model training can easily cause non-ideal decision to be made, leading to packet losses and queuing. Thus, QoE-based reward will be sparse in the RL-based CC training. The reward sparsity can slow down training or even harm the performance of the converged model. Loss-tolerant neural video codecs, on the contrary, can decode regardless of packet losses and echo the frame quality and delay back to the sender on time.

\name's reward function is defined as Equation~\ref{eq:ae-aware-rl-reward}, where $\bar q$ is a normalized frame quality of a video frame with respect to the frame's minimum and maximum possible frame quality, latency is packet latency (RTT) in ms, and $a=0.1$ is the penalty coefficient on the latency term.
\begin{equation}
    R=\sum_i (\bar{q}_i + a\cdot \text{Latency}_i) / n
    \label{eq:ae-aware-rl-reward}
\end{equation}
\mypara{Training}
The RL model is trained online against network traces and videos in a network simulator which replay network traces with a duration of 30 seconds. The network traces are randomly generated based the network parameters in Table~\ref{tab:trace-parameters}. The total number of training steps is 720000. Pre-recorded video profiles are used to simulate video codec behavior and look up frame quality of GRACE encoded frames and the profiles are uniform randomly sampled during the training process.



\section{Evaluation}
\label{sec:eval}
Our key finds are as follows:
\begin{packeditemize}
    \item \textbf{Learning efficiency:} \name running with the loss-tolerant NVC reduces the training time by 41\% compared to other prior RL-based CCs.
    \item \textbf{Better QoE:} When testing on network traces from the same training distribution, \name boosts the mean video quality by 0.3 to 1.6dB\%, lower the tail frame delay by 3 to 200ms, and reduces the video stalls by 20\% to 77\% in comparison with other baseline RTC CCs.
\end{packeditemize}

\subsection{Setup}
\label{subsec:eval-setup}
\mypara{Testbed implementation} We implement and compare different RTC CCs running with GRACE in a packet-level simulator, which replays network traces. All RTC CCs are run with padding enabled so that CCs can probe the available bandwidth efficiently. The simulator achieves the video encoding and decoding by replaying GRACE profiles which are collected on a server with 2 Nvidia A40 GPUs by profiling GRACE for various videos under diverse packet loss conditions. In this work, we only focus on how the congestion control algorithm affect the QoE and network-level performance so we assume the frame encoding and decoding is negligible. Optimizing the encoding and decoding time of the neural video codec is out of the scope of this work.

\mypara{Baselines} We compare the \name with several baselines. These baselines include human handcrafted CCs, RL-based CCs (OnRL, Aurora), and an oracle CC.

Firstly, the traditional human handcrafted CCs include GCC, Salsify, and FBRA. GCC~\cite{GCC}, a widely used CC in WebRTC applications, measures packet delay gradient and packet loss rate to detect network congestion and adjusts the target bitrate of video codecs at each frame. Salsify uses smoothed packet jitter to estimate the target sending rate and encodes a frame with two bitrates at a time to fast adapt the bandwidth fluctuation. 
FBRA sends FEC data in addition to actual video data to fast probe available bandwidth and provide protection against packet loss. 

Secondly, the RL-based baselines include OnRL and Aurora. To mimic OnRL's federated learning logic that aims to provide generalizability across diverse networks, we train OnRL across diverse network environments instead of aggregating models from different users. Unlike OnRL, Aurora is trained offline and deployed online without further finetuning.

Lastly, we add an oracle CC which knows the future bandwidth change and perfectly matches the target video bitrate to the available bandwidth. It thus serves as the optimal performance that a CC can achieve.

\mypara{Videos} Our evaluation experiments reuse the test video datasets in~\cite{Grace}, which consists of 57 videos randomly sampled from three public datasets, as shown in Table~\ref{tab:video-datasets}. Each video in the datasets is 10-30 seconds long. In this work, we use the same set of videos in both training and testing stages because we only focus on how network traces' diverse dynamics affect the training and testing of RL-based CCs.

\mypara{Network traces} The network traces used in this work include both synthetic traces and real-world collected traces. The synthetic are synthetically generated from a network trace generator used in~\cite{Genet}. Each network trace is described in five dimensions including link bandwidth, minimum link round trip time (RTT), bandwidth change interval, random packet loss rate, and queue capacity with their ranges in Table~\ref{tab:trace-parameters}. The real-world collected traces are from Pantheon dataset~\cite{pantheon}.

\mypara{Quality metrics}
To compare the performance of a RTC CC, in addition to network-level statistics, we report the QoE of a real-time video communication session across the following three aspects~\cite{Grace}.
\begin{packeditemize}
    \item {\bf Video quality} of a frame is measured by structural similarity index measure (SSIM). We report SSIM in dB, computed as $-10log(1-SSIM)$~\cite{Grace, Salsify,insitu}.
    \item {\bf Realtimeness} is measured by 98th percentile (p98) of frame delay (time gap between the frame's encoding and decoding).
    \item {\bf Smoothness} of a video is measured by video stall (an inter-frame gap exceeding 200ms~\cite{macmillan2021measuring}). Like GRACE, we report the average number of video stalls per second and the ratio of video stall time over the entire video length.
\end{packeditemize}

\begin{table}[t]
    \begin{centering}
        \begin{tabular}{ccccc}
            \hline
            {\bf Dataset} & {\bf \makecell{\# of\\videos}} & {\bf \makecell{Length\\(s)}} & {\bf Size} & {\bf Description} \\
            \hline\hline
            Kinetics & 45 & 450 & \makecell{720p\\360p} & \makecell{Human actions \\ and
            interaction \\ with objects} \\
            \hline
            Gaming & 5  & 100 & 720p & Game recordings \\
            \hline
            FVC & 7 & 140 & 1080p & \makecell{Video calls \\ (indoor/outdoor)} \\
            \hline
            \textbf{Total} & 57 & 690 &  & \\
            \hline
            \hline
        \end{tabular}
    \caption{Video datasets used.}
    \label{tab:video-datasets}
    \end{centering}
\end{table}

\begin{table}[t]
    \begin{centering}
        \begin{tabular}{cc}
            \hline
            {\bf Parameter} & {\bf Range}  \\
            \hline\hline
            Link Bandwidth (Mbps) & [0.6, 6] \\
            \hline
            Minimum link RTT (ms) & [2, 200] \\
            \hline
            Bandwidth change interval (s) & (0, 15] \\
            \hline
            Random packet loss rate & [0, 5\%] \\
            \hline
            Queue capacity (packets) & [1, 100] \\
            \hline
            \hline
        \end{tabular}
    \caption{Network parameter ranges used to generate network traces used in training and testing.}
    \label{tab:trace-parameters}
    \end{centering}
\end{table}

\subsection{Learning efficiency and asymptotic performance}
We first compare \name and prior RL-based CC, in terms of their \textit{convergence speed} and \textit{asymptotic performance} (\ie test performance over new test environments drawn independently from the training distribution).

\begin{figure}[t]
    \centering
    \includegraphics[width=0.83\columnwidth]{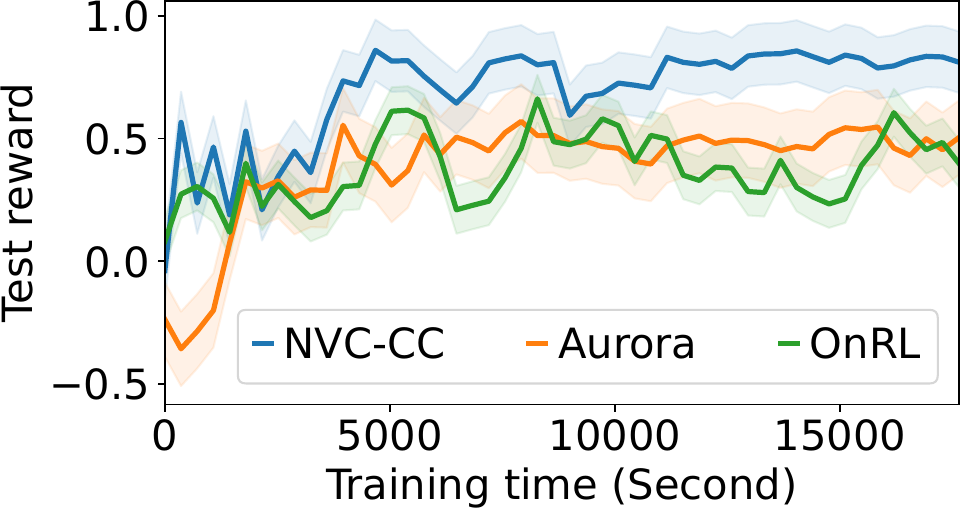}
    \tightcaption{\bf \name training ramps up faster and converges to a better test
    reward than the baseline RL-based CCs.}
    \label{fig:learning-curve}
\end{figure}

We train \name and baseline RL-based CCs across network environments generated from the distribution described in Table~\ref{tab:trace-parameters} with three different random seeds. Then we test them in new environments from the same distribution during the entire training process. Figure~\ref{fig:learning-curve} plots how the validation reward guided by GRACE (defined in Equation~\ref{eq:ae-aware-rl-reward}) of each RL-based CC changes over training time.
\name converges to the best test reward of 0.859, which is 50\% and 26\% better than that of Aurora and that of OnRL respectively.
It takes \name 78 minutes to converge to the best reward, which is approximately 41\% faster than both Aurora and OnRL.


\subsection{Performance evaluation on synthetic network traces}
\begin{figure*}[t]
    \centering
    \includegraphics[width=0.89\linewidth]{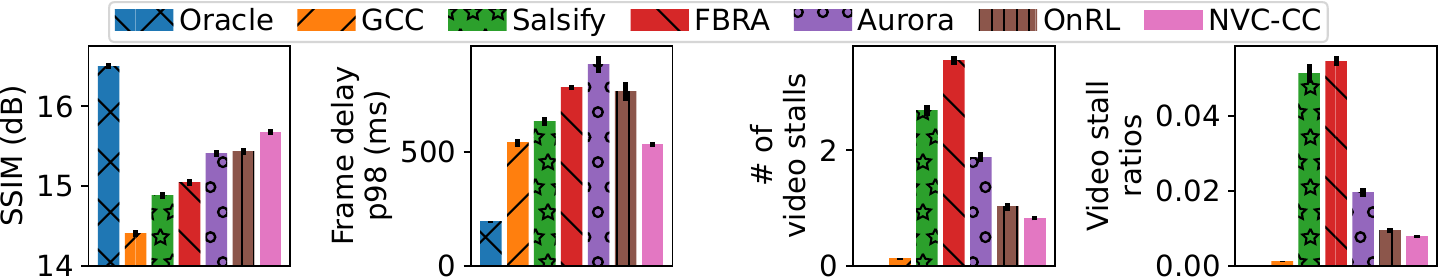}
    \tightcaption{\bf QOE breakdown when evaluating on synthetic network traces from the training distribution.}
    \label{fig:qoe-breakdown}
    \vspace{10pt}
\end{figure*}

We then compare \name and baseline CCs on the network traces which are drawn from the training distribution but unseen in the training process. The performance is broke down in terms of the real-time video communication QoE metrics mentioned in \S\ref{subsec:eval-setup}. We run them with GRACE on 2850 video sessions. These video sessions are composed by 50 network environments randomly sampled from the distribution in Table~\ref{tab:trace-parameters} and 57 videos described in Table~\ref{tab:video-datasets}.
Their QoE breakdown is shown in Figure~\ref{fig:qoe-breakdown}. \name outperforms the traditional rule-based RTC CCs by more than 1.6dB in average frame quality (SSIM). It outperforms OnRL and Aurora by at least 0.3dB. In terms of realtimeness, it leads the rule-based CCs by 3 to 200ms and leads Aurora and OnRL by 200ms. \name causes fewer video stalls than most of rule-based CCs and RL-based CCs except GCC which is known to be conservative. Salsify is outperformed by \name because GRACE is not a functional video codec Salsify requires and its fast bandwidth adaptation mechanism does not work well. FBRA is outperformed by the most of other approaches because it often overshoots the network capacity by abusing FEC data too much.

We also break down the CC's performance from the network level. From Table~\ref{tab:network-level-perf}, it is clear that \name is able to aggressively consume the available bandwidth by trading off minor degree of packet losses as it is aware of loss-tolerant NVC performance via the reward in Equation~\ref{eq:ae-aware-rl-reward}.

\begin{table}[t]
    \begin{centering}
        \begin{tabular}{cccc}
            \hline
            {\bf CC} & {\bf \makecell{Tput\\(Mbps)}} & {\bf \makecell{Delay p98\\(ms)}} & {\bf \makecell{Loss\\(\%)}} \\
            \hline
            \hline
            Oracle & 3.17 & 113 & 0.1 \\
            \hline
            GCC & 1.22 & 287 & 1.5 \\
            \hline
            Salsify & 1.72 & 272 & 5.6 \\
            \hline
            FBRA & 2.19 & 478 & 18.9 \\
            \hline
            Aurora & 2.59 & 376 & 4.6 \\
            \hline
            OnRL & 2.12 & 325 & 2.4 \\
            \hline
            \bf{\name} & 2.55 & 355 & 3.5 \\
            \hline
            \hline
        \end{tabular}
        \caption{Network-level performance breakdown.}
    \label{tab:network-level-perf}
    \end{centering}
\end{table}

\subsection{Performance evaluation on real-world network traces}
\begin{figure*}[t]
    \centering
    \begin{subfigure}[b]{\linewidth}
        \centering
        \includegraphics[width=0.89\linewidth]{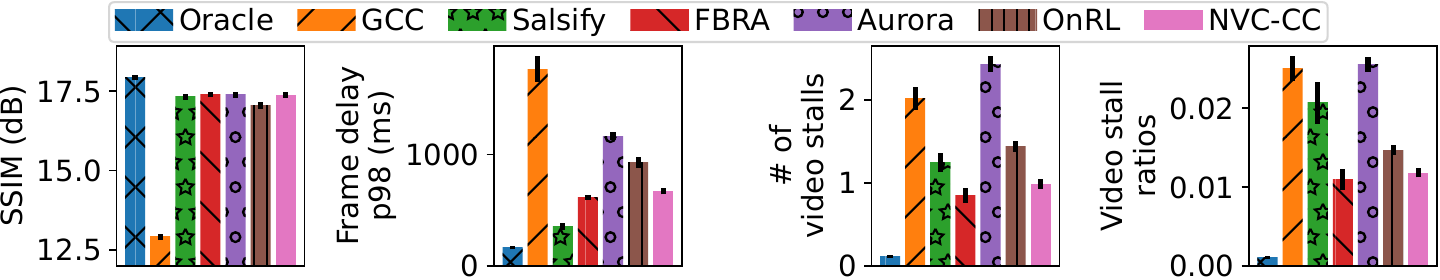}
        \caption{Ethernet}
        \label{fig:qoe-breakdown-real-trace-ethernet}
    \end{subfigure}
    \begin{subfigure}[b]{\linewidth}
        \centering
        \includegraphics[width=0.89\linewidth]{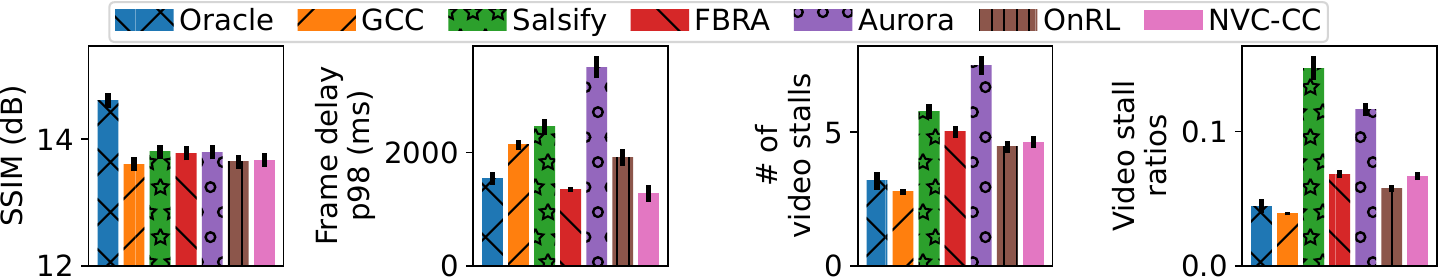}
        \caption{Cellular}
        \label{fig:qoe-breakdown-real-trace-celluar}
    \end{subfigure}
    \tightcaption{\bf QOE breakdown when evaluating on real-world network traces.}
    \label{fig:qoe-breakdown-real-trace}
    \vspace{8pt}
\end{figure*}

In addition to evaluating the CCs on the network traces from the training distribution, we test \name's generalizability on network traces outside the training distribution by evaluating them real-world collected traces from Pantheon dataset~\cite{pantheon}. These traces contain two major network settings--ethernet and cellular and they were collected on 10 different links among nodes globally. Each CC is the tested on 570 video sessions and the resulted QoE breakdown are shown in Figure~\ref{fig:qoe-breakdown-real-trace-ethernet} and Figure~\ref{fig:qoe-breakdown-real-trace-celluar} respectively. Although \name still outperforms prior RL-based CCs along all four QoE metrics in both network settings, its benefits over traditional rule-based CCs are marginal, indicating that \name's generalizability on network traces outside its training distribution still need to improve.

\section{Related work}
\mypara{RTC congestion control} Unlike congestion controls~\cite{tcpcubic, tcpvegas, tcpcc} which are designed for reliability and in order delivery in bulky data transfer scenario, RTC congestion controls are specifically designed for packet delivery timeliness by leveraging implicit congestion indicators like delay gradient ~\cite{GCC, Salsify} and explicit congestion signals ~\cite{SCReAM} and they are adapted for various network environments ~\cite{NADA}.

Starting from Remy~\cite{Remy}, machine learning models, especially RL, are used to automatically create more adaptive controlling logics beyond the hand-crafted heuristics in the field of congestion control and rate adaptation. Some \cite{Remy, Aurora, RFEC, Mamba, Loki, Pensieve, AlphaRTC, Concerto} explore pure RL solutions while others~\cite{OnRL, Orca, Genet} employ a combination of RL model and hand-crafted heuristics.
But all these prior works are specifically designed to work with traditional video codec which cannot decode an incomplete frame or video codec agnostic. There still does not exist an RTC CC optimized for video codecs that tolerate frame-level packet loss.

\mypara{Loss-resilient techniques}
Before the emergence of NVCs, traditional loss-resilience techniques provide the protection for real time coding. One major category is forward error correction (FEC). FEC (\ie Reed-Solomon codes~\cite{ReedSolomonCode}, LDPC~\cite{NearShannonCode} fountain code~\cite{FountainCode}, streaming codes~\cite{StreamingCode, Tambur}, and reinforcement learning-based code~\cite{rlcode}) estimates and inserts data redundancy before the data is sent to the network. FEC fails to recover an incomplete video frame or leads to bandwidth waste if the frame-level packet loss rate is underestimated or overestimated.

The other major category is postprocessing error concealment (EC) which manipulates encoding (\ie INTRA-mode macroblock ordering~\cite{intromacroblockencoding}, slide interleaving~\cite{sliceinterleaving}, and flexible macroblock ordering ~\cite{macroblockordering}) guarantee the encoded packets to be decodable when they are partially received. The decoder reconstructs lost data based on the received data, using classic heuristics~\cite{wang2013novel, zhou2010efficient, kumar2006error} or neural-network-based inpainting~\cite{kang2022error, sankisa2018video, mathieu2015deep}. Due to the encoder’s lack of awareness of the decoder’s postprocessing, each encoded packet contains limited redundancy and the reconstructed video quality drops faster than NVCs when packet loss rate increases.

In this work, we focus on exploiting neural video codec properties only and treat congestion controls working with traditional loss-resilient techniques as baselines.

\mypara{Safe reinforcement learning} To avoid online action exploration, the ``learning offline, running online" strategy trains RL models offline in a simulator or emulator and then deploys the trained model in real networks. Although RL models achieve good performance during the training stage, the performance is unsatisfactory in real network environments~\cite{mao2008real}. There are other safe reinforcement learning techniques: using human handcrafted safeguard policies to recover the system from risky state ~\cite{OnRL, mao2019towards}; converting a human handcrafted policy into a black-box neural network and then fusing with RL model~\cite{Loki}; prioritizing rewarding network environments in training ~\cite{Genet} or even adversarial training~\cite{robustifying}. However, none of the prior works exploit the properties of video codecs to achieve safety in RL-based solutions for the field of networking.






\section{Discussion}
\label{sec:discuss}
While the preliminary results show early promise of \name, there are still limitations in its methodology and future steps needed to be done.

\mypara{Video and codec dependency}
Because \name's reward used in training relies on the frame quality of a video frame, there naturally exists a dependency between the trained \name model and the video codec as well as the training videos. The dependency on video codec indicates that a different \name model should be used when the loss-tolerant NVC is updated. To address the dependency on training videos, more research effort needs to be invested in measuring and improving \name RL model's generalization over videos with diverse content.

\mypara{Sim-to-real generalization}
Similar to other RL models trained in a simulated environment and deployed in the real world, \name also faces the simulation to real world generalization problem. Even though \name is trained using online RL training, our experiments are conducted by replaying network traces in network simulator. Thus, an important future step to investigate is to measure and improve \name's generalizabilty over real-world network environments.

\mypara{Fairness} On the Internet, many different congestion control protocols need to interact, and ours can possibly behave unfairly in some scenarios. For example, being aware of loss-tolerant NVC allows \name to be aggressive in increasing packet sending rate and unaffected by minor packet drops. It is possible that \name takes up the majority of link capacity and forces traditional TCP to back off quickly.

\mypara{Scalability}
Due to the reliance on the performance of loss-tolerant NVCs, \name also inherits some limitations from NVCs. For instance, GRACE is not optimized enough to run at 30 fps on resource-constrained devices that barely sustain classic video codec. Thus, \name cannot be trained to support 30 fps real time streaming on such devices. Another example is that GRACE focuses on unicast video communication instead of multi-party video conferencing. Therefore, the performance of \name in multi-party video conferencing scenario is still unknown.

\section{Conclusion}
\label{sec:conclusion}
We present \name, an RL-based congestion control algorithm for real time video communication applications. For the first time, we analyze the tradeoff between quality of experience during RL training stage and RL model convergence speed. We reveal that the commonly used safeguard policies used to prevent risky actions and recover the networked system to safe states can cause sampling efficiency and slow down the RL model convergence. \name addresses the training efficiency problem by training RL model on top of loss-tolerant neural video codec without the need of safeguard policies. Our evaluation shows that \name's training with loss-tolerant neural video codec improves RL training efficiency and achieves better quality of experience in various network environments and video workloads.

\noindent \textbf{Ethics:} This work does not raise any ethical issues.

%
\begin{acks}
The project is funded by NSF CNS-2146496, CNS-2131826, CNS-2313190, CNS-1901466, and UChicago CERES Center.
\end{acks}

\bibliographystyle{ACM-Reference-Format}
\bibliography{reference}

\end{document}